\documentclass{article}

\usepackage{arxiv}

\usepackage[utf8]{inputenc} 
\usepackage[T1]{fontenc}    
\usepackage{hyperref}       
\usepackage{url}            
\usepackage{booktabs}       
\usepackage{amsfonts}       
\usepackage{nicefrac}       
\usepackage{microtype}      
\usepackage{lipsum}
\usepackage{graphicx}
\usepackage[bottom]{footmisc}
\usepackage[utf8]{inputenc}
\usepackage[T1]{fontenc}
\usepackage{multirow}
\usepackage[english]{babel}

\usepackage{hyperref, url, booktabs, amsfonts, nicefrac, microtype, lipsum, comment}

\usepackage{verbatim}
\usepackage{cite}
\usepackage{amsmath,amssymb,amsfonts}
\usepackage{algorithmic}
\usepackage{multicol, blindtext}
\usepackage{textcomp}
\usepackage[table,xcdraw]{xcolor}
\def\BibTeX{{\rm B\kern-.05em{\sc i\kern-.025em b}\kern-.08em
    T\kern-.1667em\lower.7ex\hbox{E}\kern-.125emX}}

\usepackage{multirow}
\usepackage{xr-hyper}
\usepackage{hyperref, url}
\usepackage[normalem]{ulem}
\usepackage{pdfpages}

\makeatletter
\newcommand*{\addFileDependency}[1]{
  \typeout{(#1)}
  \@addtofilelist{#1}
  \IfFileExists{#1}{}{\typeout{No file #1.}}
}
\makeatother

\newcommand*{\myexternaldocument}[1]{%
    \externaldocument{#1}%
    \addFileDependency{#1.tex}%
    \addFileDependency{#1.aux}%
}
\myexternaldocument{Supplement}

\title{Improved supervised prediction of aging-related genes via weighted dynamic network analysis}

\author{
  Qi Li$\dagger$, Khalique Newaz$\dagger$, and Tijana Milenkovi\'{c}*\\
  Department of Computer Science, Center for Network \& Data Science, and Eck Institute for Global Health\\
  University of Notre Dame\\
  Notre Dame, IN 46556, USA \\
  \texttt{\{qli8, knewaz, tmilenko\}@nd.edu} \\
  $\dagger$: equal contribution; *: corresponding author.
}

\begin{document}
\maketitle

\begin{abstract}
\noindent\textbf{Background:} This study focuses on the task of supervised prediction of aging-related genes from -omics data. Unlike gene expression methods for this task that capture aging-specific information but \emph{ignore interactions} between genes (i.e., their protein products), or protein-protein interaction (PPI) network methods for this task that account for PPIs but the PPIs are \emph{context-unspecific}, we recently integrated the two data types into an aging-specific PPI subnetwork, which yielded more accurate aging-related gene predictions. However, a \emph{dynamic} aging-specific subnetwork did not improve prediction performance compared to a \emph{static} aging-specific subnetwork, despite the aging process being dynamic. This could be because the dynamic subnetwork was inferred using a naive Induced subgraph approach. Instead, we recently inferred a dynamic aging-specific subnetwork using a methodologically more advanced notion of network propagation (NP), which improved upon Induced dynamic aging-specific subnetwork in a different task, that of \emph{unsupervised} analyses of the aging process.\\
\noindent\textbf{Results:} Here, we evaluate whether our existing NP-based dynamic subnetwork will improve upon the dynamic as well as static subnetwork constructed by the Induced approach in the considered task of \emph{supervised} prediction of aging-related genes. The existing NP-based subnetwork is unweighted, i.e., it gives equal importance to each of the aging-specific PPIs. Because accounting for aging-specific edge weights might be important, we additionally propose a \emph{weighted} NP-based dynamic aging-specific subnetwork. We demonstrate that a predictive machine learning model trained and tested on the weighted subnetwork yields higher accuracy when predicting aging-related genes than predictive models run on the existing unweighted dynamic or static subnetworks, regardless of whether the existing subnetworks were inferred using NP or the Induced approach.\\
\noindent\textbf{Conclusions:} Our proposed weighted dynamic aging-specific subnetwork and its corresponding predictive model could guide with higher confidence than the existing data and models the discovery of novel aging-related gene candidates for future wet lab validation.\\
\end{abstract}

\section{Introduction} \label{sect:intro}
\subsection{Motivation and related work}\label{sect:intro_motivation}

Incidence of many complex diseases, such as diabetes, cancer, osteoarthritis, cardiovascular, and Alzheimer's disease, increases with age \cite{campisi2013aging}. Even recent and widespread COVID-19 is highly related to aging. Namely, according to the United States Centers for Disease Control and Prevention\footnote{\url{https://www.cdc.gov/nchs/nvss/vsrr/COVID19}}, as of March 24, 2021, ${\sim}$81\% of all deaths related to COVID-19 occurred in the age range of 65 years and above.

Understanding the molecular mechanisms behind the aging process, including comprehensive and accurate identification of human genes implicated in aging, is important for studying and treating such aging-related diseases \cite{bolignano2014aging, fabris2017review}. However, analyzing human aging via wet lab experiments is difficult due to long human life span and ethical constraints \cite{faisal2014dynamic,faisal2014global}. Analyzing human aging computationally can fill this gap. This includes identification (i.e., prediction) of aging-related genes via supervised learning from human -omics data \cite{freitas2011data, kerepesi2018prediction, fabris2017review}, which is the task that we focus on in this paper. By supervised prediction of aging-related genes, we mean that a part of current aging-related knowledge is used when predicting which genes are linked to aging, and the remaining part of that knowledge is used when evaluating the predictions. On the other hand, an orthogonal and thus complementary task, which is outside of the scope of this paper, is that of unsupervised prediction of aging-related genes. By this, we mean that no current aging-related knowledge is used when making predictions, and instead, all of the aging-related knowledge is used only to evaluate the predictions.

Approaches for our task at hand -- supervised prediction of aging-related genes -- can be categorized as follows: \emph{(i)} those that use gene expression data alone, \emph{(ii)} those that use protein-protein interaction (PPI) network data alone, and \emph{(iii)} those that combine gene expression data with PPI network data.

Approaches from the first category predict a gene as aging-related if its expression level varies with age \cite{jia2018analysis, lu2004gene, berchtold2008gene, holtman2015induction, simpson2011microarray}. While such approaches do capture aging-specific information, they ignore interactions between genes, i.e., their protein products. This is their drawback, because proteins  carry out cellular functioning, including the aging process, by interacting with each other \cite{kirkwood2005understanding}. 

Approaches from the second category predict a gene as aging-related if its position (i.e., node representation/embedding/feature) in the PPI network is ``similar enough'' to the network positions of known aging-related genes \cite{freitas2011data, fang2013classifying, fabris2016extensive, kerepesi2018prediction}. While these approaches do consider PPIs that carry out cellular functioning, their drawback is that the PPIs are context-unspecific, i.e., the PPIs span different conditions, such as cell types, tissues, diseases, environments, or patients. In the context of our study, this means that the PPIs are not aging-specific, i.e., they span different ages.

Approaches from the third category address the above drawbacks by considering both aging-specific gene expression data and context-unspecific PPI network data. Earlier approaches of this type extracted genes' features from each of gene expression data and PPI network data individually and then concatenated the features \cite{freitas2011data,kerepesi2018prediction}. As such, they integrated the features rather than the data. Consequently, they still considered the context-unspecific PPI network. More recent efforts, including by our group, first integrated the two data types to infer an aging-specific subnetwork of the entire context-unspecific PPI network, and then analyzed the resulting subnetwork \cite{faisal2014dynamic,Elhesha2019}. Specifically, these studies used a traditional \emph{Induced approach}: given gene expression data for different ages and an entire context-unspecific PPI network (which happens to be \emph{static}), the studies formed a subnetwork snapshot for each age, where each snapshot consisted of all genes that were significantly expressed (i.e., active) at the given age and all PPIs from the entire network that exist between the active genes. That is, they formed a given age-specific snapshot by taking the Induced subgraph on the active genes at that age. Then, they combined snapshots for all ages into a \emph{dynamic} aging-specific PPI subnetwork. Such a subnetwork, being dynamic, is meant to capture how network positions of genes change with age.  

These studies used their inferred dynamic subnetworks in \emph{unsupervised} aging-related tasks \cite{faisal2014dynamic,Elhesha2019}. So, while their inferred subnetworks are relevant for this paper, their unsupervised analyses of the subnetworks fall outside of its scope.

On the other hand, \emph{supervised} analyses from our other, even more recent work \cite{li2019supervisedBIBM,li2020supervisedTCBB} \emph{are} directly relevant for this paper. In that work, we performed supervised prediction of aging-related genes from three networks: \emph{(i)} the dynamic aging-specific subnetwork inferred by Faisal and Milenkovi\'{c} \cite{faisal2014dynamic}, \emph{(ii)} a static (still aging-specific) counterpart of this dynamic subnetwork, and \emph{(iii)} the entire (also static) context-unspecific PPI network from which the above dynamic aging-specific subnetwork was inferred. Note that the static Induced subnetwork from point (ii) was constructed by aggregating all nodes and edges over all snapshots of the dynamic Induced subnetwork. Then, first, we examined whether using the dynamic or static aging-specific subnetwork improved the supervised prediction accuracy compared to using the entire (static) context-unspecific PPI network. Indeed, we found this to hold overall (i.e., it held for many although not all of the considered evaluation measures). Second, because the aging process is dynamic, we examined whether using the dynamic aging-specific subnetwork was superior to using the static aging-specific subnetwork. Surprisingly, we found this \emph{not} to hold overall (i.e., it did not hold for most of the considered evaluation measures).

This unexpected finding could be because the dynamic subnetwork was inferred using the Induced approach, which is quite naive as it considers \emph{all} PPIs from the context-unspecific network  that exist between \emph{only} the active genes at a given age. However, first, not all PPIs between the active genes might be equally ``important'', and the Induced approach has no way of identifying  the most important of all such PPIs \cite{newaz2018improving}. Second, the Induced approach fails to consider any inactive genes that might critically connect the active genes in the network \cite{newaz2018improving}.

Recently, an alternative, methodologically more advanced notion of \emph{network propagation (NP)} was used for inference of a \emph{static} context-specific subnetwork in the context of \emph{cancer} \cite{cowen2017network}. NP maps expression levels (i.e., activities) onto the genes in the entire context-unspecific PPI network. Then, NP propagates the activities via random walk or diffusion, to assign condition-specific weights to the nodes (genes, i.e., their protein products) or edges (PPIs) in the entire PPI network. Finally, NP assumes that it is the highest-weighted network regions that are the most relevant for the condition of interest, i.e., such regions form the context-specific subnetwork. Hence, as opposed to the Induced approach, first, NP assigns weights to PPIs that can help identify the most ``important'' PPIs. Second, NP can consider a non-active gene if, for example, the gene is connected to sufficiently many active genes.

Our group extended two prominent NP approaches, i.e., NetWalk \cite{Komurov2010} and HotNet2 \cite{leiserson2015pan}, to allow for the inference of \emph{dynamic} context-specific subnetworks in the context of \emph{aging} \cite{newaz2018improving}. Then, we evaluated the NP-based aging-specific subnetworks in an \emph{unsupervised} learning context, finding that NP-based dynamic aging-specific subnetworks outperform the dynamic aging-specific subnetwork created using the Induced approach \cite{newaz2018improving}. In particular, a dynamic aging-specific subnetwork inferred using the NetWalk method outperformed every other considered dynamic aging-specific subnetwork \cite{newaz2018improving}. This is why in this paper, out of all existing NP-based aging-specific subnetworks, we focus on the NetWalk-based one. For methodological details of NetWalk and HotNet2, please refer to their original publications \cite{Komurov2010,leiserson2015pan} or to our work on unsupervised study of aging \cite{newaz2018improving}. Note that while the inferred NP-based subnetworks are relevant for this paper, the unsupervised analyses of the networks by Newaz and Milenkovi\'{c} \cite{newaz2018improving} fall outside of its scope.

With the above information in mind, in this paper, a key question that we aim to answer is whether using an NP-based dynamic aging-specific subnetwork (either the existing one by NetWalk or a novel one that we propose as a contribution of our study; see below) will outperform every other available network (i.e., the dynamic and static aging-specific subnetworks resulting from the Induced approach, static counterparts of the dynamic NP-based aging-specific subnetworks, and the entire static context-unspecific PPI network). This question has several subquestions, as discussed next.

\subsection{Our study and contributions}\label{sect:intro_ourstudy}

Our first subquestion is whether using an NP-based dynamic aging-specific subnetwork improves the \emph{supervised} prediction of aging-related genes compared to using any of the existing (dynamic or static) Induced aging-specific subnetworks. To answer this, we consider two NP-based dynamic subnetworks and two Induced subnetworks, as follows. 

One of the considered NP-based dynamic aging-specific subnetworks, which we refer to as \emph{NetWalk-Dynamic}, was inferred in our existing study \cite{newaz2018improving} (see Section \ref{sect:method-data-subnet}). This subnetwork was constructed by using NP (specifically, NetWalk) to propagate expression activities onto the genes in the entire context-unspecific PPI network in order to assign aging-specific weights to the PPIs. However, NetWalk does not have a procedure of extracting a context-specific (i.e., aging-specific) subnetwork from such a weighted PPI network \cite{newaz2018improving}. Therefore, in our previous study \cite{newaz2018improving}, we first applied a weight threshold and then treated all PPIs with weights above the threshold as aging-specific. Then, we extracted the aging-specific PPIs (and proteins involved in these PPIs) as the resulting aging-specific subnetwork. Consequently, this subnetwork is unweighted, as the PPIs were assigned equal importance (see \cite{newaz2018improving} for more detail). This approach, which was the best one could do at the time, suffers from two limitations. First, it is hard to choose a meaningful PPI weight threshold, because there is no clear guideline about which threshold is meaningful. So, one needs to empirically evaluate subnetworks constructed at multiple thresholds, which is computationally inefficient. Second, this approach cannot distinguish between different aging-specific PPIs, because it assumes that each such PPI carries the same amount of aging-specific information as the others. To address these limitations, in this study, we aim to infer a novel, i.e., \emph{weighted} NP-based dynamic aging-specific subnetwork, with hope that accounting for PPI weights, i.e., for aging-specific importance of the PPIs, will perform better in our task of supervised prediction of aging-related genes. We refer to this weighted subnetwork as \emph{$w$NetWalk-Dynamic} (see Section \ref{sect:method-data-subnet}). We aim to evaluate whether at least one of NetWalk-Dynamic and $w$NetWalk-Dynamic is better than the dynamic and static Induced subnetworks from our previous studies that have already been discussed above \cite{li2019supervisedBIBM, li2020supervisedTCBB}. We refer to the latter two as Induced-Dynamic and Induced-Static, respectively.

Assuming that NP performs better than the Induced approach, our second subquestion is whether at least one of the two NP-based dynamic aging-specific subnetworks is better than static counterparts of both of the NP-based dynamic subnetworks. We do this because aging is a dynamic process, and so dynamic network analysis of aging should outperform static network analysis of aging. To answer this, for reasons described in Section \ref{sect:method-data-subnet}, we construct two distinct static counterparts for NetWalk-Dynamic, referred to as NetWalk-Static and NetWalk-Static*, respectively. We also construct a (single) static counterpart for $w$NetWalk-Dynamic, referred to as $w$NetWalk-Static*. 

Recall that we hypothesize that \emph{weighted} NP will be better than \emph{unweighted} NP, which is why we infer $w$NetWalk-Dynamic in the first place. So, our third subquestion is to test this hypothesis -- whether $w$NetWalk-Dynamic performs better than NetWalk-Dynamic. Note that for as comprehensive evaluation as possible, we also consider the entire context-unspecific network that all aging-specific subnetworks are inferred from, with expectation that the latter will outperform the entire network. We refer to the entire network as \emph{Entire}. We summarize the eight considered (sub)networks in Table \ref{table:nettype}.

\begin{table}[!htp]
    \centering
    \caption{Summary of the eight considered (sub)networks. All but Entire are aging-specific subnetworks.}
    \label{table:nettype}
    \begin{tabular}{|c|c|c|c|c|}
    \hline
    Network name  & Weighted & Unweighted & Dynamic & Static  \\ \hline
    Entire         &    &  \checkmark  & & \checkmark    \\ \hline
    Induced-Dynamic  &  & \checkmark  &   \checkmark  &       \\ \hline
    Induced-Static & & \checkmark  &    & \checkmark      \\ \hline
    NetWalk-Dynamic  &  & \checkmark  &   \checkmark  &       \\ \hline
    NetWalk-Static & & \checkmark  &    & \checkmark      \\ \hline
    NetWalk-Static* & & \checkmark  &    & \checkmark      \\ \hline
    $w$NetWalk-Dynamic  & \checkmark &   &   \checkmark  &       \\ \hline
    $w$NetWalk-Static* & \checkmark &  &    & \checkmark      \\ \hline
\end{tabular}
\end{table}

For each (sub)network, we develop multiple supervised predictive models where each model consists of a \emph{(i)} feature, \emph{(ii)} feature dimensionality reduction choice, and \emph{(iii)} classifier. \emph{(i)} Regarding features, given that our (sub)networks are of different types (unweighted vs. weighted, and static vs. dynamic), we need to choose features that fit a given network type. For unweighted dynamic and static (sub)networks, we use existing sophisticated features from our previous studies \cite{li2019supervisedBIBM, li2020supervisedTCBB}. For weighted dynamic subnetwork, we consider five existing features \cite{opsahl2010, Barrat2004, anderson1962}. However, because the existing weighted features are simple extensions of the notion of unweighted network centrality, we believe that we can do better. So, we propose a new way to extract weighted dynamic features of nodes, with hope that they will outperform the existing weighted dynamic features. Thus, our fourth subquestion is whether using our proposed weighted dynamic features will outperform the existing weighted dynamic features on $w$NetWalk-Dynamic. In the process, as a control, we propose static counterparts of our new weighted dynamic features to evaluate whether the latter will be superior, as we would expect. \emph{(ii)} Regarding dimensionality reduction choices, for each feature, we consider its full version and its dimensionality-reduced version. \emph{(iii)} Regarding classifiers, we couple each feature (version) with three prominent classifiers. For details, see Section \ref{sect:methods_predmodel}.

For each predictive model, we evaluate its performance via cross-validation \cite{li2020supervisedTCBB}: we train on a subset of known aging- and non-aging-related genes and test on the remaining subset of known aging- and non-aging-related genes. To define known aging- and non-aging-related gene labels for classification, we rely on highly confident ground truth data, i.e., GenAge \cite{tacutu2017human}. We evaluate performance of a predictive model in terms of the area under the precision-recall (AUPR) curve, precision, recall, and F-score. Given all predictive models of a (sub)network, we choose the best predictive model that yields the highest AUPR score. Then, we compare the different (sub)networks, each under its selected best model, via all four prediction accuracy measures. 

Because cancer is known to be aging-related \cite{campisi2013aging}, in addition to cross-validation, we use cancer-related data to validate the performance of each (sub)network. Specifically, given a (sub)network, we measure the number of cancer-related genes among its novel gene predictions, i.e.,  genes predicted as aging-related that are currently not associated with aging.

In summary, the above four subquestions can be seen as testing the following four hypotheses, \emph{(i)} whether using an \textbf{NP-based} aging-specific subnetwork will outperform all \textbf{Induced} aging-specific subnetworks, \emph{(ii)} whether using an \textbf{NP-based dynamic} aging-specific subnetwork will outperform all \textbf{NP-based static} aging-specific subnetworks, \emph{(iii)} whether using the \textbf{weighted dynamic} NP-based aging-specific subnetwork will outperform the \textbf{unweighted dynamic} NP-based aging-specific subnetworks, and \emph{(iv)} given the weighted dynamic subnetwork, whether using our \textbf{proposed} features will outperform \textbf{existing} features.

We find that overall, all of our four hypotheses hold with respect to both cross-validation and cancer-related validation. Most importantly, our proposed weighted dynamic aging-specific subnetwork (i.e., $w$NetWalk-Dynamic) along with our proposed weighted dynamic feature performs the best among all considered (sub)networks. This finding justifies the need of using a weighted dynamic aging-specific subnetwork in the task of predicting aging-related genes, which is the key contribution of our study.

Because $w$NetWalk-Dynamic performs the best, we believe that it has potential to guide future experimental validation of its aging-related gene predictions that are currently not known to be aging-related. As the first attempt to this, we identify all six such genes that are predicted only by $w$NetWalk-Dynamic. We manage to validate all of the six genes by manually searching for a link between them and the aging process in PubMed articles. This finding is another contribution of our study. 


\section{Results and discussion}\label{sect:results}

We present the results for when using the GenAge-based definition of aging- and non-aging-related gene labels in the first six subsections. Human genes in GenAge are sequence orthologs of aging-related genes in model species. Because not all human genes have sequence orthologs in model species \cite{Mazza2009,khalturin2009more}, using GenAge may miss aspects of the aging process that are unique to human \cite{danchin2018bacteria}. So, we consider another source of aging-related knowledge obtained by studying human directly (rather than doing so indirectly from model species), namely the down-regulated aging-related genes from genotype-tissue expression (GTEx) project \cite{jia2018analysis}, i.e., GTEx-DAG. We discuss the findings when using the GTEx-DAG-based label definition in Section \ref{sect:results_DAG}. 

\subsection{The best predictive model for each (sub)network} 

In this subsection, we comment on how we select the best predictive model for each (sub)network. Recall that a predictive model consists of three components, i.e., a node feature, a feature dimensionality reduction choice, and a classifier. Regarding node features, we consider four types of features, each corresponding to one of the four network types (i.e., each feature can only be extracted from its corresponding network type). We summarize the considered features in Table \ref{tab:features}. Regarding dimensionality choices, for each feature, we consider its full as well as principal component analysis (PCA) reduced versions. Regarding classifiers, for each feature version, we run three classifiers, i.e., logistic regression (LR), naive Bayes (NB), and support vector machine (SVM) with radial basis function (rbf) (SVM-rbf). For methodological details and reasons of why we use these components, see Section \ref{sect:Methods}.

\begin{table}[!htp]
    \centering
    \caption{The considered node features and their types, i.e., weighted dynamic, weighted static, unweighted dynamic, and unweighted static. Features in bold are (the best versions of) our proposed features, and the rest are the considered existing features. For the discussion on how we select the best of the existing weighted dynamic features and the best of the proposed weighted dynamic features, see Supplementary Section \ref{ssect:best_feature} and Supplementary Fig. \ref{sfig:best_feature}.}
    \label{tab:features}
    \begin{tabular}{l|l|l|}
    \cline{2-3} & Weighted    & Unweighted   \\ \hline
    \multicolumn{1}{|l|}{Dynamic} &  \begin{tabular}[c]{@{}l@{}}\textbf{Diff-nobin-2}  \\ DegC-wt,   ClusC-wt, CloseC-wt, BetwC-wt, EigenC-wt\end{tabular} & \begin{tabular}[c]{@{}l@{}}DGDV, GoT, GDC, \\ ECC, KC, DegC, CentraMV\end{tabular} \\ \hline
    \multicolumn{1}{|l|}{Static}  & \textbf{Static-nobin-2} &  SGDV, UniNet, 30BPIs   \\ \hline
\end{tabular}
\end{table}

Recall that for each of the eight (sub)networks, we select the best predictive model that yields the highest AUPR score. Briefly, the eight selected models (each corresponding to a (sub)network) encompass five features, two dimensionality choices, and three classifiers (Table \ref{table:selected-model}). While some individual components of a model are somewhat preferred over others, there is no model as a whole that consistently works the best over all types of (sub)networks. So, in summary, there is no clear pattern regarding which component of which model works well on which (sub)network. This inconsistency of components for the selected models across all (sub)networks stresses out the need for our comprehensive evaluation of running all of the different models, in order to give each (sub)network the best-case advantage. Also, it emphasizes the importance of a shift in the field of machine learning development of interpretable predictive models, in order to allow for understanding the effect of each model component on the prediction outcome. 

\begin{table}[!htp]
    \centering
    \caption{The selected best predictive model for each (sub)network with respect to the GenAge-based aging- and non-aging-related gene labels. The model is represented as ``X+Y+Z'', where X represents the selected feature, Y represents the selected dimensionality choice (i.e., None or PCA), and Z represents the selected classifier. For dimensionality choices, ``None'' means the selected feature is in its full version, and ``PCA'' means the selected feature is in its PCA-reduced version.}
   \label{table:selected-model}
    \begin{tabular}{ccc}
    \hline
    Entire  & Induced-Dynamic   & Induced-Static  \\ 
    SGDV + None + NB    & DGDV + None + NB  & SGDV + None + NB   \\ \hline
    NetWalk-Dynamic &  NetWalk-Static  & NetWalk-Static* \\
     DGDV + None + NB &  30BPIs + None + SVM-rbf & 30BPIs + PCA + LR  \\ \hline
    $w$NetWalk-Dynamic  & $w$NetWalk-Static* &\\ 
    Diff-nobin-2 + None + LR  & Static-nobin-2 + None + LR & \\ \hline
\end{tabular}
\end{table}

\subsection{Using NP-based vs. Induced aging-specific subnetworks}

\begin{figure}[!ht]
    \centering
    \includegraphics[width=0.7\linewidth]{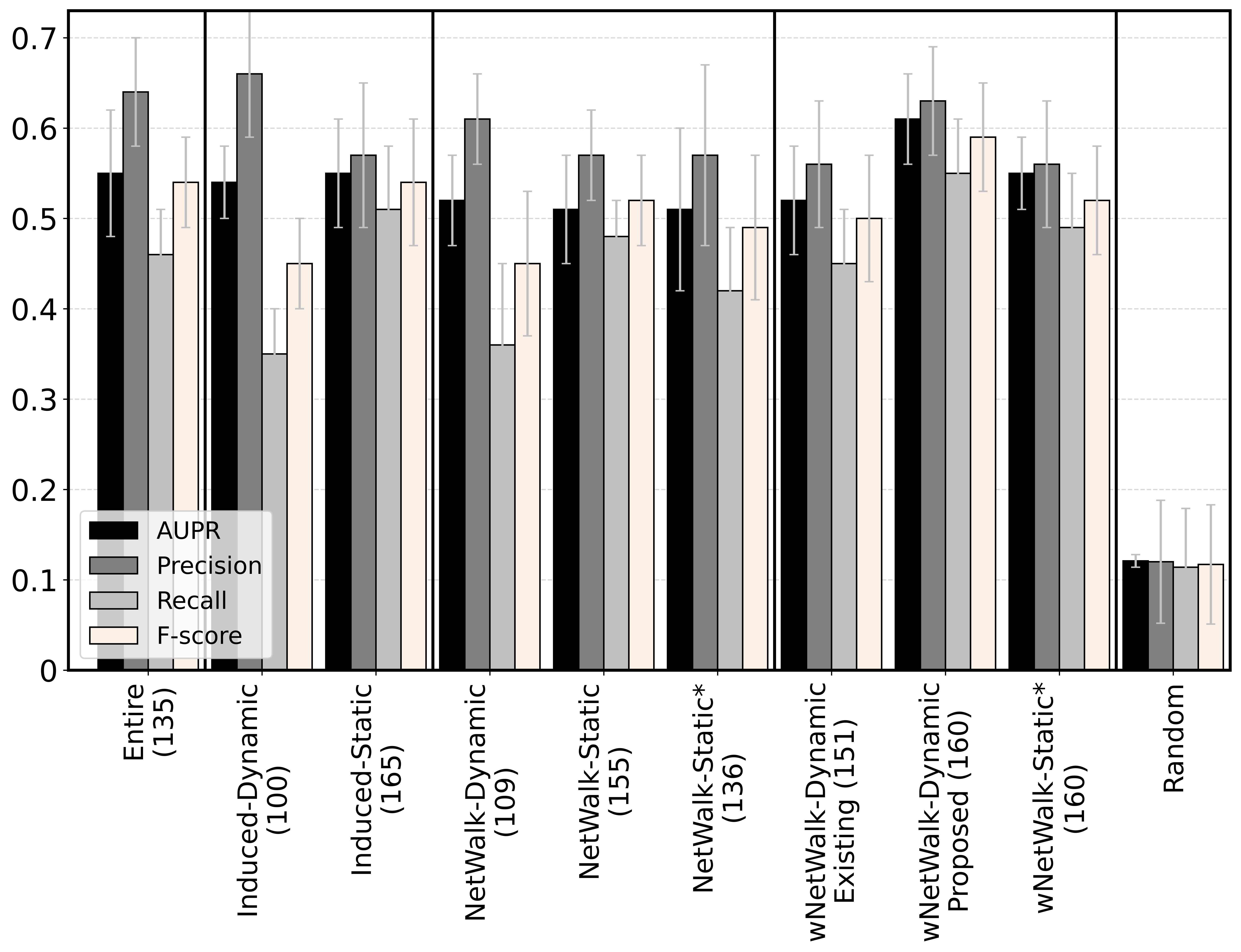}
    \caption{Prediction accuracy in terms of AUPR, precision, recall, and F-score for the eight (sub)networks, each under its best predictive model, when using the primary GenAge-based definition of aging- and non-aging-related gene labels. On the $x$-axis, the number in the parenthesis after the name of each (sub)network corresponds to the number of predictions. Note that we have two groups of results for $w$NetWalk-Dynamic, i.e., $w$NetWalk-Dynamic Existing and $w$NetWalk-Dynamic Proposed. The former corresponds to using the best existing weighted dynamic feature on $w$NetWalk-Dynamic, and the latter corresponds to using the best proposed weighted dynamic feature on $w$NetWalk-Dynamic.}
    \label{fig:APRF}
\end{figure}

\begin{figure}[!ht]
    \centering
    \includegraphics[width=0.6\linewidth]{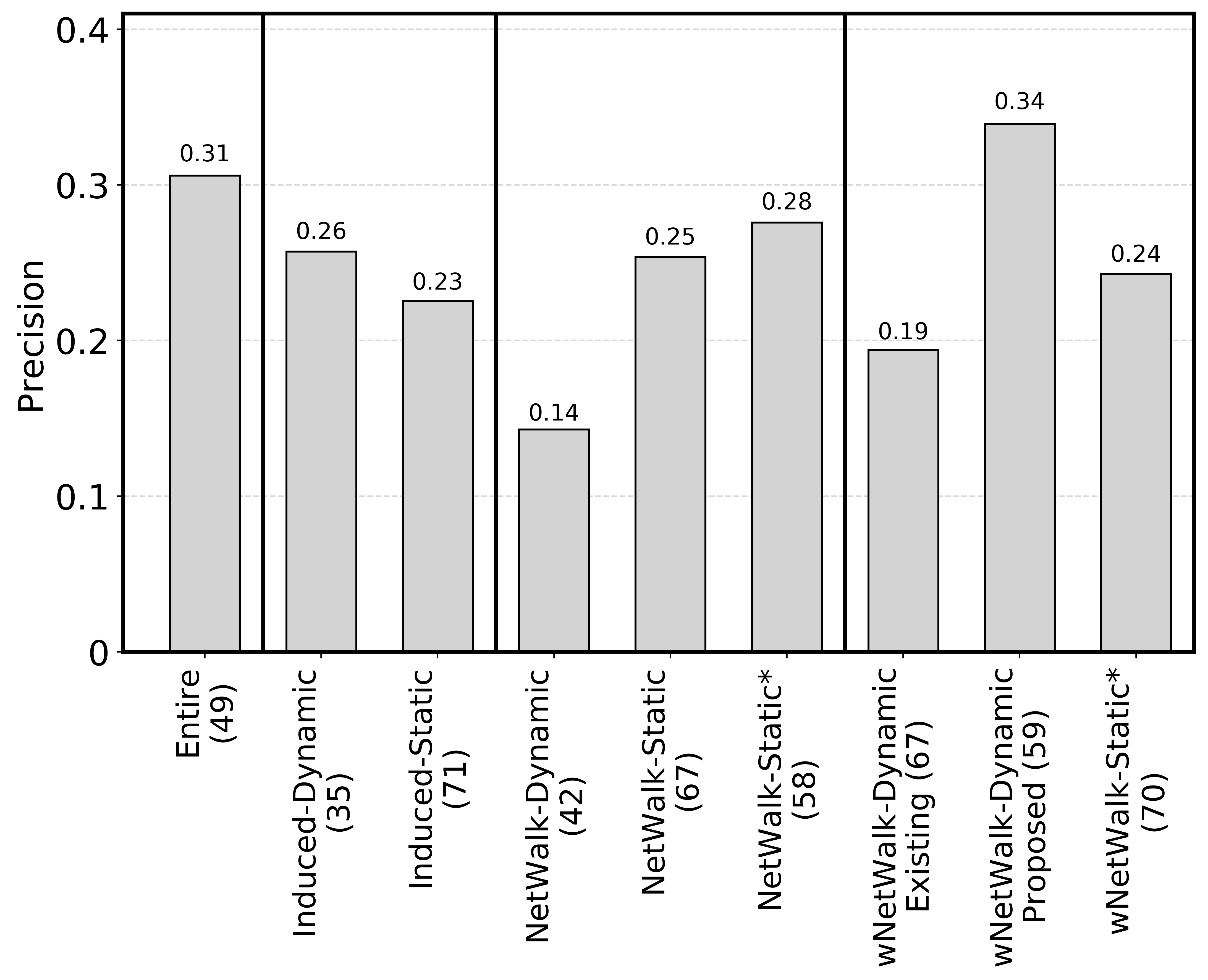}
    \caption{Cancer-related validation of the novel aging-related gene predictions for the eight considered (sub)networks, each under its best predictive model, when using the primary GenAge-based definition of aging- and non-aging-related gene labels. On the $x$-axis, the number in the parentheses after the name of each (sub)network is the number of the novel predictions. The precision score of a (sub)network is the percentage of novel predictions that are validated by the cancer-related data. Note again that we have two groups of results for $w$NetWalk-Dynamic, i.e., $w$NetWalk-Dynamic Existing and $w$NetWalk-Dynamic Proposed, which we already explained in the caption of Fig. \ref{fig:APRF}.}
    \label{fig:cancer_val}
\end{figure}

We first comment on our hypothesis \emph{(i)}: whether using an \textbf{NP-based} aging-specific subnetwork will outperform all \textbf{Induced} aging-specific subnetworks. We begin with testing whether the best of the existing (unweighted) dynamic subnetworks from our previous study \cite{newaz2018improving} (i.e., NetWalk-Dynamic) and its two (unweighted) static counterparts (i.e., NetWalk-Static and NetWalk-Static*) outperforms the two Induced subnetworks (i.e., Induced-Dynamic and Induced-Static). However, we find this not to be the case. \underline{In the context of cross-validation}, NetWalk-Dynamic, which performs the best among the above three unweighted NP-based subnetworks, has a tied performance to Induced-Dynamic in terms of recall and F-score but has a lower performance than Induced-Dynamic in terms of AUPR and precision (Fig. \ref{fig:APRF}). Also, NetWalk-Dynamic has a lower performance than Induced-Static in terms of AUPR, recall, and F-score, but a marginally better performance than Induced-Static in terms of precision. Similarly, \underline{in the context of cancer-related validation}, NetWalk-Dynamic has a lower performance than both Induced-Dynamic and Induced-Static (Fig. \ref{fig:cancer_val}).

These results match our postulation that unweighted NetWalk-based subnetworks could be suboptimal because all of their PPIs are treated with equal importance (see Section \ref{sect:intro}). This is exactly why we propose to infer a weighted subnetwork using NetWalk in the first place. Indeed, when we look into whether the best of the weighted NP-based subnetworks, i.e., $w$NetWalk-Dynamic, outperforms all Induced subnetworks, we find this to be the case in terms of most of the considered evaluation measures, as follows. 

\underline{In the context of cross-validation}, $w$NetWalk-Dynamic, when analyzed via our proposed novel weighted dynamic features, outperforms all Induced subnetworks in terms of AUPR, recall, and F-score. In terms of precision, while $w$NetWalk-Dynamic is marginally (i.e., not statistically significant) inferior to Induced-Dynamic, it still outperforms Induced-Static. The superiority of $w$NetWalk-Dynamic is statistically significant when compared to Induced-Dynamic in terms of AUPR, recall, and F-score, and when compared to Induced-Static in terms of AUPR and F-score (Fig. \ref{fig:APRF}) (adjusted $p$-values $\le 0.05$). 

\begin{figure}[!ht]
    \centering
    \includegraphics[width=0.7\linewidth]{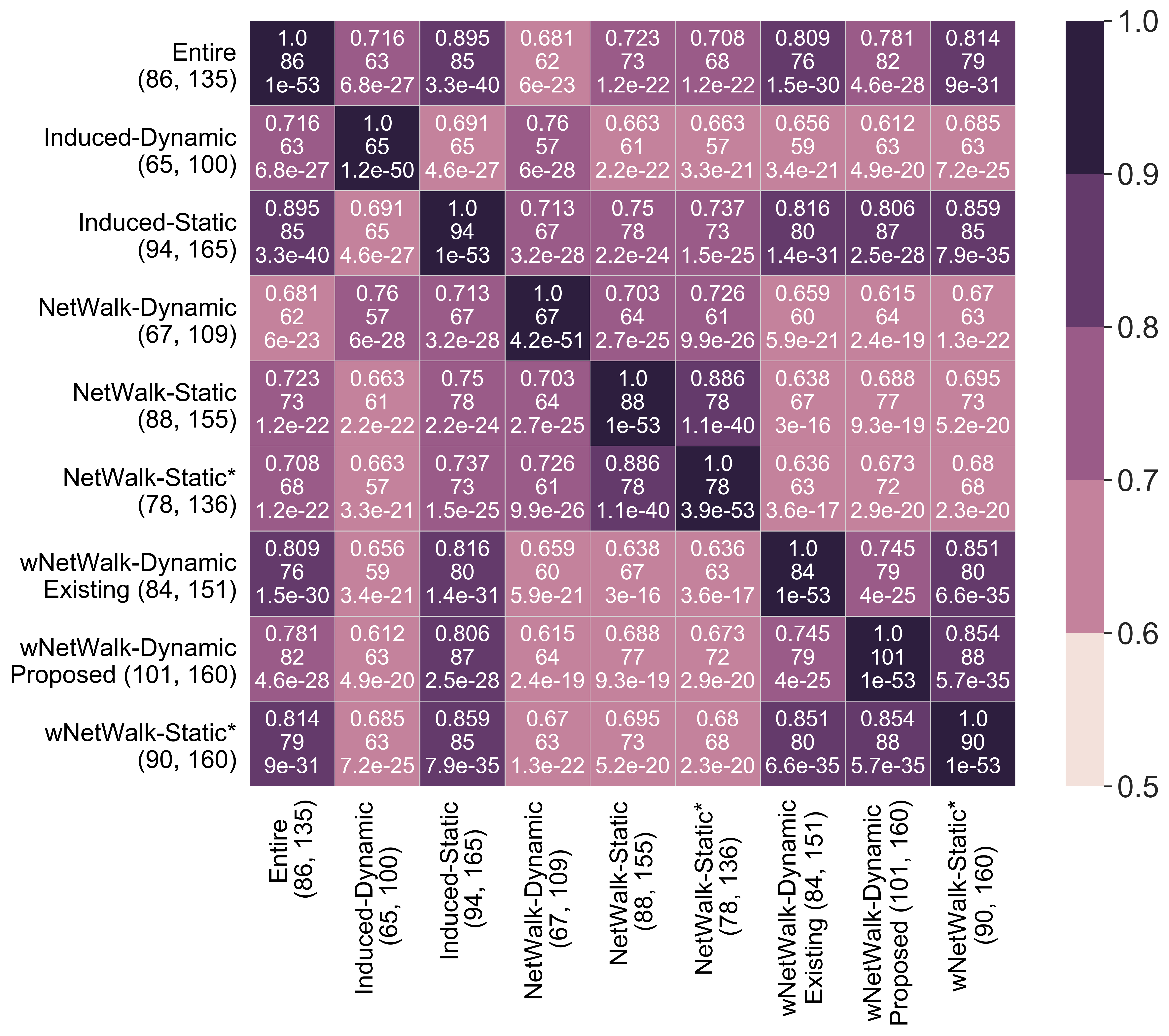}
    \caption{Overlaps (expressed as Jaccard indices) between true positive predictions of the eight considered (sub)networks, each under its best predictive model, when using the primary GenAge-based definition of aging- and non-aging-related gene labels. On the $x$- and $y$-axes, the numbers in the parentheses after the name of each subnetwork are the number of true positive predictions (i.e., known aging-related genes) and the number of total predictions. Within each table/matrix cell, the number on the first line is the Jaccard index of the overlap between the corresponding pair of subnetworks; the number on the second line is the raw overlap size, i.e., the actual count of true positives that are in the overlap between the two subnetworks; the number on the third line is the adjusted $p$-value of the overlap with respect to the hypergeometric test. The darker the color of a given cell, the larger the Jaccard index value, i.e., the higher the overlap. Note that we have two groups of results for $w$NetWalk-Dynamic, i.e., $w$NetWalk-Dynamic Existing and $w$NetWalk-Dynamic Proposed, which we already explained in the caption of Fig. \ref{fig:APRF}.}
    \label{fig:pred_overlap}
\end{figure}

For all four cross-validation prediction accuracy measures, all (sub)networks perform statistically significantly better than the random approach (adjusted $p$-values $\le 0.032$). That is, each (sub)network predicts a majority (over 56\%) of the genes that are labeled as aging-related (i.e., are known aging-related genes). So, we expect the pairwise overlaps of all (sub)networks' true positives to be reasonably large. Indeed, we find that overlaps are statistically significantly high for all (sub)network pairs. The largest overlap over all (sub)network pairs is 89.5\%, the smallest one is 61.2\%, and the average one is 72.9\% (Fig. \ref{fig:pred_overlap}). We note that even though our $w$NetWalk-Dynamic marginally underperforms Induced-Dynamic in terms of precision, $w$NetWalk-Dynamic not only correctly predicts 63 out of all 65 known aging-related genes that Induced-Dynamic predicts (i.e., 96.9\% of them), but it also \emph{correctly} predicts 38 known aging-related genes that Induced-Dynamic \emph{does not} predict. Similarly, we examine pairwise overlaps of the (sub)networks' predicted genes that are currently labeled as non-aging-related, i.e., of their \emph{novel} aging-related gene predictions. We find that all of the pairwise overlaps are statistically significantly high. The largest overlap over all pairs is 62.2\%, the smallest one is 17.9\%, and the average one is 33.9\% (Supplementary Fig. \ref{sfig:pred_overlap_NP}).

\underline{In the context of cancer-related validation} (Fig. \ref{fig:cancer_val}), we again find that $w$NetWalk-Dynamic (under our proposed novel weighted dynamic features) performs better than all Induced aging-specific subnetworks. That is, 34\% of novel aging-related gene predictions (i.e., predicted genes currently not known to be aging-related) made by $w$NetWalk-Dynamic are present in the cancer-related data, while the best Induced aging-specific subnetwork (i.e., Induced-Dynamic) only gets 26\% of novel predictions validated by the cancer-related data.

All of the above results signify that our hypothesis \emph{(i)}, i.e., that using an NP-based aging-specific subnetwork is better than using all Induced aging-specific subnetworks, holds with respect to almost all evaluation measures, i.e., AUPR, recall, F-score, pairwise prediction overlaps, as well as the cancer-related validation. This result stresses out the significance of proposing  our $w$NetWalk-Dynamic.

\subsection{Using dynamic vs. static NP-based  aging-specific subnetworks}

Here, we evaluate our hypothesis \emph{(ii)}: whether using the best of the \textbf{NP-based dynamic} aging-specific subnetwork will outperform all \textbf{NP-based static} aging-specific subnetworks. Because $w$NetWalk-Dynamic is superior to (unweighted) NetWalk-Dynamic, it is $w$NetWalk-Dynamic that we compare to the NP-based static subnetworks (i.e., NetWalk-Static, NetWalk-Static*, and $w$NetWalk-Static*). 

\underline{In the context of cross-validation}, we find that $w$NetWalk-Dynamic indeed outperforms all three static NP-based subnetworks in terms of all four prediction accuracy measures (Fig. \ref{fig:APRF}). The superiority of $w$NetWalk-Dynamic is statistically significant when compared to NetWalk-Static* in terms of recall, and when compared to $w$NetWalk-Static* in terms of AUPR (adjusted p-values $\le 0.05$), while all other superiorities of $w$NetWalk-Dynamic are non-significant. When we look into the pairwise prediction overlaps between the dynamic vs. static NP-based subnetworks (Fig. \ref{fig:pred_overlap}), we confirm  $w$NetWalk-Dynamic's superiority. Namely, it is $w$NetWalk-Dynamic that not only has the highest precision, but also has the highest number of predictions than any of the other NP-based subnetworks.

\underline{In the context of cancer-related validation}, we again confirm the superiority of $w$NetWalk-Dynamic (Fig. \ref{fig:cancer_val}). That is, 34\% of the novel aging-related genes predicted by $w$NetWalk-Dynamic are present in the cancer-related data. At the same time, the best static NP-based  subnetwork (i.e., NetWalk-Static*) only has 28\% of its novel predictions present in the cancer-related data. 

Therefore, hypothesis \emph{(ii)}, i.e., that using an \textbf{NP-based dynamic} aging-specific subnetwork is better than using all \textbf{NP-based static} subnetworks, holds with respect to all evaluation measures. This result further stresses out the significance of proposing our weighted dynamic NP-based aging-specific subnetwork.

\subsection{Using weighted vs. unweighted dynamic NP-based aging-specific subnetworks}

Here, we evaluate our hypothesis \emph{(iii)}: whether using the \textbf{weighted dynamic} NP-based aging-specific subnetwork (i.e., $w$NetWalk-Dynamic) is better than using the \textbf{unweighted dynamic} NP-based aging-specific subnetwork (i.e., NetWalk-Dynamic). Indeed, we find this to be the case. 

\underline{In the context of cross-validation}, as shown in Fig. \ref{fig:APRF}, $w$NetWalk-Dynamic outperforms NetWalk-Dynamic in terms of all four prediction accuracy measures. Of the four, the superiority is statistically significantly high  (adjusted p-values $\le 0.05$) in terms of recall and F-score. When we look into the pairwise prediction overlaps between $w$NetWalk-Dynamic and  NetWalk-Dynamic, we find that $w$NetWalk-Dynamic not only correctly predicts 64 out of all 67 known aging-related genes that NetWalk-Dynamic predicts (i.e., 95.5\% of them), but it also \emph{correctly} predicts 37 known aging-related genes that NetWalk-Dynamic \emph{does not} predict.

\underline{In the cancer-related validation}, we again find $w$NetWalk-Dynamic to be better (Fig. \ref{fig:cancer_val}). Specifically, among the novel aging-related genes (currently not known to be aging-related) predicted by $w$NetWalk-Dynamic and NetWalk-Dynamic, 34\% and only 14\%, respectively, are among the cancer-related genes.

Henceforth, hypothesis \emph{(iii)}, i.e., that using $w$NetWalk-Dynamic is superior to using NetWalk-Dynamic, holds with respect to all evaluation measures. This result again stresses out the importance of inferring our weighted NP-based dynamic aging-specific subnetwork.

\subsection{Using our proposed vs. existing weighted dynamic features}

Here, we evaluate our hypothesis \emph{(iv)}: whether using \textbf{our proposed} weighted dynamic features will be better than using the \textbf{existing} weighted dynamic features. To validate our hypothesis \emph{(iv)}, given $w$NetWalk-Dynamic, it would suffice for our best weighted dynamic feature to outperform the best existing weighted dynamic feature. So, given $w$NetWalk-Dynamic, we choose the best predictive model based on the considered existing weighted dynamic features and the best predictive model based on our proposed weighted dynamic features, (see Section \ref{sect:methods_predmodel} and Supplementary Section \ref{ssect:best_feature}). Then, we compare the two chosen best models using cross-validation, prediction overlap, and cancer-related validation.

\underline{In the context of cross-validation}, we find that our best proposed feature outperforms the best existing feature in terms of all four prediction accuracy measures. Of the four measures, the superiority is statistically significant in terms of AUPR, recall, and F-score (i.e., adjusted p-values $\le 0.05$) (Fig. \ref{fig:APRF}). When we look into the pairwise prediction overlaps between our best proposed feature and the best existing feature, we find that the best proposed feature not only correctly predicts 79 out of all 84 known aging-related genes that the best existing feature predicts (i.e., 94.0\% of them), but it also \emph{correctly} predicts 22 known aging-related genes that the best existing feature \emph{does not} predict.

\underline{In the context of cancer-related validation}, our best proposed feature again wins over the best existing feature. Specifically, 34\% of the novel aging-related genes predicted by the former and only 19\% of the novel aging-related genes predicted by the latter are present in the cancer-related data.

Therefore, our hypothesis \emph{(iv)}, i.e., that using the best of our proposed weighted dynamic features is better than using all of the existing weighted dynamic features, holds with respect to all evaluation measures.

\subsection{Literature validation of novel aging-related gene predictions resulting from the best subnetwork, i.e., $w$NetWalk-Dynamic}

Because $w$NetWalk-Dynamic (under the best of our proposed features) is the best subnetwork among all eight considered (sub)networks, those genes that are predicted by $w$NetWalk-Dynamic as aging-related but are currently not known to be linked to aging (i.e., that are novel aging-related gene predictions) are the most confident predictions for future experimental validation. There are 59 such novel predictions. Because it would be extremely time consuming to validate all of the 59 novel predictions, we aim to validate using literature search those novel predictions that are predicted \emph{only by} $w$NetWalk-Dynamic but not by any other (sub)network. There are six such predictions, i.e., ATF1, GLI3, JUNB, NRP1, SMC1A, and SPI1. When we manually search for a direct (or indirect) link between these six genes and the aging process (or an aging-related disease) in PubMed articles, we manage to validate all six of them. Namely:

\begin{itemize}
    \item \underline{ATF1} was shown to be related to the pathogenesis of the Alzheimer's disease \cite{ding2019construction}, which is known to be highly related to aging \cite{goedert2006century}. 
    \item \underline{GLI3} was shown to be linked to tumorigenicity of the colorectal cancer \cite{iwasaki2013hedgehog}, which was shown to be mainly affecting the elderly \cite{holt2009colon}. 
    \item \underline{JUNB}'s expression levels were shown to increase in all immune cells during the aging process \cite{zheng2020human}.
    \item \underline{NRP1} was shown to be overexpressed in various human carcinoma cell lines and primary tumors \cite{beck2011vascular}, which are the main causes of skin cancer that usually appeared in old people \cite{garcovich2017skin}.
    \item The mutation of \underline{SMC1A} was detected in colorectal cancer \cite{sarogni2019overexpression},  which was shown to be mainly affecting the elderly \cite{holt2009colon}.
    \item \underline{SPI1} was identified as being upregulated in Alzheimer's disease \cite{srinivasan2020alzheimer}, which is known to be highly related to aging \cite{goedert2006century}. 
\end{itemize}

\subsection{Results when using our secondary aging-related gene label definition with respect to GTEx-DAG} \label{sect:results_DAG}

All results reported so far are for when using the GenAge-based labels to define genes as aging- vs. non-aging-related. Here, we report results when using the GTEx-DAG-based labels. The (sub)networks’ best predictive models when using GTEx-DAG are shown in Supplementary Table \ref{stable:selected-model}. 

\underline{In the context of cross-validation}, like for the GenAge-based labels, we do see that $w$NetWalk-Dynamic again performs the best among all tested eight (sub)networks in terms of all prediction accuracy measures. Unlike for the GenAge-based labels, we now also see that (unweighted) NetWalk-Dynamic outperforms both Induced subnetworks. These results further support the need for NP-based weighted dynamic aging-specific subnetworks (Supplementary Fig. \ref{sfig:APRF-DAG}).

However, in our previous studies \cite{li2020supervisedTCBB}, we already showed that for the two Induced subnetworks, prediction accuracy is not nearly as good under the GTEx-DAG-based gene labels as it is under the GenAge-based gene labels. Here, we confirm that the same holds for the NP-based (weighted and unweighted) subnetworks. Their prediction accuracy scores are not statistically significantly more accurate than the random approach (adjusted p-values $\ge 0.05$).

Because of the random-like performance of all (sub)networks with respect to GTEx-DAG, we do not perform analysis of pairwise prediction overlaps. Also, we do not make any novel predictions from the GTEx-DAG-based classifiers. Consequently, the cancer-related validation cannot be performed. The inferior performance of GTEx-DAG compared to GenAge is surprising for two reasons. First, genes in GTEx-DAG have been linked to aging directly in human, while those in GenAge have been linked to aging indirectly based on their sequence similarities to aging-related genes in model species. So, GTEx-DAG should better capture aspects of the aging process that are unique to human. Given this, and given that we have aimed to extract human aging-related knowledge from human PPI subnetworks, one would expect GTEx-DAG to perform better. However, it does not. This could be because GenAge may be more accurate than GTEx-DAG. Second, the considered (sub)networks have been formed by integrating gene expression data with PPI data. Given that GTEx-DAG is also gene expression-based, while GenAge is sequence-based, one would expect GTEx-DAG to perform better. However, it does not. This could be because GTEx-DAG and GenAge are highly complementary (only 17 genes are labeled as aging-related in both), and because our predictive models may be better suited for the aging-related knowledge present in GenAge.


\section{Conclusions} \label{sect:conclusion}

In this study, we have proposed a novel NP-based weighted dynamic aging-specific subnetwork, along with novel weighted dynamic node features, which combined yield better prediction accuracy in supervised prediction of aging-related genes. We have used NP in hope of improving upon Induced that has been used traditionally. Indeed, this is what we observe. 

Our findings are not quantitatively robust to the choice of ground truth aging-related data.
Namely, our predictive models work better for sequence-based aging-related data from model species than for gene expression-based aging-related data from human. Development of predictive models that will be capable of capturing well human-unique aspects of the aging process (which do not exist in model species) is important. 

A way to improve upon the current findings, including potentially on their robustness to data choice, could be as follows. First, in the current study, we use traditional machine learning where the user has to define a feature first and then ``plug-in'' this feature into an off-the-shelf classifier. However, lots of advancements have been made in deep learning on graphs, including graph convolutional networks (GCNs) for dynamic network analysis, in domains such as social or citation networks \cite{pareja2019evolvegcn,manessi2020dynamic}. So, it might be worth it to adopt or customize these existing GCNs to the computational biology domain and specifically to the task of studying the aging process. Second, in the current study, the aging-specific (sub)networks are inferred from the entire HPRD human PPI network that was curated a while ago. Recently, a newer version of human PPI network data named HuRI \cite{luck2020reference} was curated. So, it might be worth it to adopt the newer human PPI network, which is the subject of our future work. 
 
We have studied human aging from PPI data. Nonetheless, our work can be applied to other types of biological networks, other species, or other biological processes, such as disease progression.


\section{Methods}\label{sect:Methods}

\subsection{Data \label{sect:method-data}}
\subsubsection{The considered entire context-unspecific PPI network\label{sect:method-data-entirePPI}} 

The considered unweighted entire context-unspecific PPI network in this study is obtained from the HPRD database \cite{keshava2008human}. We take the largest connected component that encompasses 8,938 nodes and 35,900 edges, and we denote it as Entire. We use this network because it is the network that all of our aging-specific subnetworks (see below for details) are inferred from. 

\subsubsection{Seven considered aging-specific subnetworks\label{sect:method-data-subnet}} 

Other than Entire, we also consider seven aging-specific subnetworks that are inferred from Entire via two network inference methods, i.e., the Induced approach and the NP-based approach called NetWalk, as follows. 

First, we use the dynamic aging-specific subnetwork inferred by Faisal and Milenkovi\'{c} \cite{faisal2014dynamic} using the Induced approach. Specifically, given human gene expression data at 37 ages spanning between 20 and 99 years \cite{berchtold2008gene}, a dynamic aging-specific subnetwork consisting of 37 subnetwork snapshots, one snapshot per age, was constructed. Each subnetwork snapshot corresponded to the Induced subgraph of those genes that are significantly expressed (i.e., active) at the given age. Because the Induced approach does not map gene activities on to genes or PPIs (i.e., it does not assign weights to genes or PPIs), the 37 Induced subnetwork snapshots are unweighted. We refer the resulting Induced unweighted dynamic aging-specific subnetwork, as Induced-Dynamic. Moreover, one of our key question is whether aging-related genes can be predicted more accurately from a dynamic network than from a static network. So, we create an \emph{unweighted} static counterpart for Induced-Dynamic, as follows. We aggregate (i.e., take the union of) the nodes and edges over all 37 snapshots such that a PPI is kept if it is present in any of the 37 subnetwork snapshots. We denote this static subnetwork as Induced-Static. This allows for a fair comparison between a dynamic subnetwork against its static counterpart, as the two contain the same nodes and edges.

Other than Induced-Dynamic, we also use an NP-based dynamic aging-specific subnetwork inferred in our previous work \cite{newaz2018improving}. Key differences between the Induced approach and NP are as follows. The Induced approach considers \emph{all PPIs} from Entire that exist between \emph{only active genes}. On the other hand, with an assumption that not all PPIs between the active genes might be equally important, NP's weighting mechanism allows it to select only highly-weighted of all PPIs from Entire. Also, with a different assumption that it is important to consider both active genes and non-active genes that critically connect the active genes in Entire, NP propagates activities of highly expressed (i.e., active) genes to other (i.e., non-active) genes in the network. As such, it could include a non-active gene into its aging-specific subnetwork if the non-active gene is (directly or indirectly) linked to sufficiently many active genes. 

We extended two prominent NP approaches originally proposed for inference of a static cancer-specific subnetwork, i.e., NetWalk and HotNet2, to the problem of inferring a dynamic aging-specific subnetwork \cite{newaz2018improving}. Specifically, given an NP approach, we created a dynamic aging-specific subnetwork with 37 age-specific subnetwork snapshots, as follows. Given Entire and expression levels for genes at a given age, we used the NP approach to assign age-specific weights to every PPI of Entire, such that the higher the weight of a PPI, the more important the PPI is. We did this for each of the 37 ages and obtained 37 age-specific weighted subnetwork snapshots. Finally, to obtain an unweighted dynamic aging-specific subnetwork, we only kept those ``highly'' weighted PPIs in each of the 37 age-specific weighted subnetwork snapshots, which resulted in the most relevant aging-specific dynamic subnetwork (i.e., the collection of the 37 age-specific subnetwork snapshots).  See Newaz and Milenkovi\'{c} \cite{newaz2018improving} for details. We evaluated  dynamic aging-specific subnetworks produced by NetWalk against those produced by HotNet2, and found the former to be better, i.e., to overlap more with aging-related ground truth data \cite{newaz2018improving}. Consequently, in this paper, we only focus on the dynamic aging-specific subnetworks inferred using NetWalk. For this NP approach, we considered two versions, referred to as option $1$ and option $2$ \cite{newaz2018improving}. The key difference between these two versions lies in the different ways in which they process gene expression levels for each gene type (i.e., active or non-active genes) prior to propagating them through the network. Intuitively, the former assigns actual expression levels to only those genes that are significantly expressed (active) at a given age and ``non-informative'' (i.e., ``dummy'') expression levels to all other genes in the entire network, while the latter assigns actual expression levels to both active and non-active genes. We have shown that the dynamic subnetwork inferred using option $2$ (which was referred to as NetWalk* in the original publication \cite{newaz2018improving}) overlapped better with the aging-related ground truth data than the dynamic subnetwork inferred using option $1$ \cite{newaz2018improving}. Hence, we use the dynamic subnetwork inferred using NetWalk option $2$, which we refer to as NetWalk-Dynamic.

NetWalk-Dynamic was constructed by only keeping ``highly'' weighted PPIs. This means that a weight ``threshold'' was defined before the ``highly'' weighted PPIs were selected. In other words, those PPIs with weights less than a predefined threshold were removed. However, identifying such a threshold can be a time-consuming task. Additionally, we believe that keeping all of the PPIs and distinguishing between them using aging-specific weights is likely to capture more aging-specific information than only keeping a set of most ``important'' PPIs. That is, we aim to construct a dynamic aging-specific subnetwork such that \emph{(i)} the aging-specific information can be distinguished via the PPIs' weights and \emph{(ii)} a predefined weight ``threshold'' is not needed, as follows.

Given Entire and the gene expression data for 37 ages \cite{berchtold2008gene}, we first use NetWalk (i.e., option 2) to obtain 37 weighted subnetwork snapshots in the same manner as done by Newaz and Milenkovi\'{c} \cite{newaz2018improving} (also described above). However, unlike Newaz and Milenkovi\'{c} \cite{newaz2018improving}, we do not remove any of the PPIs from any of the 37 weighted subnetwork snapshots. Hence, each of the 37 snapshots have all of the PPIs from Entire. Second, to make the weights of PPIs comparable across different snapshots, we normalize the PPI weights over all 37 snapshots between the values of $0.01$ and $1$. Additionally, we believe that studying changes in the age-specific weights of PPIs across subsequent ages can capture more aging-specific information than just analyzing ``raw'' age-specific weights of PPIs. Hence, third, given the 37 snapshots with normalized PPI weights, we create 36 ``differential'' snapshots, as follows. Given two consecutive snapshots, i.e., $i$ and $i+1$, we create a differential snapshot, such that, for a given PPI (e.g., $Y$), we define its differential weight ($Y^{wt}_{i,i+1}$) as the percentage change from its weight in snapshot $i$ (i.e., $Y^{wt}_i$) to its weight in snapshot $i+1$ (i.e., $Y^{wt}_{i+1}$). Formally, we define the weight of the PPI in a differential snapshot ($i,i+1$) as $Y^{wt}_{i,i+1} = \frac{[Y^{wt}_{i+1} - Y^{wt}_i]\times 100}{[Y^{wt}_{i+1} + Y^{wt}_i]}$. We use the collection of 36 differential weighted subnetwork snapshots as the weighted NetWalk-Dynamic aging-specific subnetwork, which we refer to as $w$NetWalk-Dynamic.

Hence, we consider two versions of NetWalk-based dynamic aging-specific subnetworks, i.e., the \emph{unweighted} dynamic aging-specific subnetwork resulting from thresholding (NetWalk-Dynamic), and the \emph{weighted} dynamic aging-specific network without thresholding ($w$NetWalk-Dynamic). Additionally, we consider a counterpart of $w$NetWalk-Dynamic, which we obtain as follows. We follow the same procedure that we do for $w$NetWalk-Dynamic up to the step where we obtain 37 snapshots with normalized edge weights. We consider this weighted dynamic subnetwork as the ``non-differential'' counterpart of $w$NetWalk.

We consider this ``non-differential'' counterpart of $w$NetWalk because of the following reasons. $w$NetWalk-Dynamic is a ``differential'' weighted subnetwork, where the weight of an edge captures a change between two consecutive snapshots. However, the considered existing weighted dynamic features were originally not defined for a differential weighted dynamic subnetwork. To give the existing weighted dynamic features the best-case advantage, we evaluate them not only on $w$NetWalk-Dynamic but also on the non-differential counterpart of $w$NetWalk-Dynamic. Because we find that the existing weighted dynamic features perform better with respect to the non-differential counterpart of $w$NetWalk-Dynamic than with respect to $w$NetWalk-Dynamic, we only report results of the existing weighted features based on the non-differential counterpart of $w$NetWalk-Dynamic.

Similar to Induced-Dynamic, we also need static counterparts for the dynamic NP-based subnetworks. We consider two versions of static subnetworks, i.e., static and static*. Because the first version is meant for unweighted dynamic subnetworks, we create the static counterpart only for NetWalk-Dynamic, which we denote as NetWalk-Static. On the other hand, because NP was originally proposed for static context-specific subnetwork, for an NP-based dynamic subnetwork, we use NP to infer another static subnetwork counterpart, i.e., static*, as follows. We first aggregate gene expression data over all ages (so that a node is considered active if it is active at any one or more of the 37 ages) and then integrate the aggregated gene expression-based activity data with Entire using NetWalk. The resulting static aging-specific subnetwork (without thresholding) is the counterpart of $w$NetWalk-Dynamic, and is denoted as \emph{$w$NetWalk-Static*}. Then, to create the static counterpart of NetWalk-Dynamic, given $w$NetWalk-Static*, we choose the edge weight threshold that results in a static subnetwork matching the number of edges in NetWalk-Static, and we denote such static subnetwork as NetWalk-Static*. This way, the two versions of the static subnetworks, i.e., NetWalk-Static and NetWalk-Static*, are fairly comparable to each other. 

In total, we analyze Entire which is unweighted, two unweighted Induced subnetworks, three unweighted NP-based subnetworks, and two weighted NP-based subnetworks. That is, we consider $1 + 2 + 3 + 2 = 8$ (sub)networks (Table \ref{table:netsize}).

\begin{table}[!ht]
    \centering
    \caption{The sizes of the eight considered (sub)networks. The size of a dynamic subnetwork is the average over its 37 (or 36, depending on whether the considered dynamic subnetwork has 37 or 36 snapshots) subnetwork snapshots. The two numbers delimited by ``;'' are node and edge counts of the corresponding (sub)network.}
    \label{table:netsize}
        \begin{tabular}{|c|c|c|c|c|}
        \hline
        Entire  & Induced-Dynamic & Induced-Static & NetWalk-Dynamic                \\ \hline
        8,938; 35,900  & 4,696; 15,062   & 6,371; 22,975  & 3,749; 9,617            \\ \hline
        NetWalk-Static & NetWalk-Static* & $w$NetWalk-Dynamic  & $w$NetWalk-Static* \\ \hline
        4,973; 14,509  & 5065; 14,509    & 8,938; 35,900       & 8,938; 35,900      \\ \hline
    \end{tabular}
\end{table}

\subsubsection{Six considered aging-related ground truth data\label{sect:method-groundtruth}}

The purpose of using the following data is described in the next Section.

\begin{itemize}
    \item GenAge \cite{tacutu2017human} is a highly trusted source of human aging-related genes, most of which are sequence orthologs of experimentally validated aging-related genes in model organisms. Of all 307 human genes from GenAge, 185 are present in all eight considered (sub)networks. Henceforth, we denote the 185-gene set as \underline{GenAge}.
    
    \item  GTEx \cite{jia2018analysis}  contains 863 genes whose expressions decrease with age (i.e., are \emph{down-regulated}), of which 347 are present in all eight considered (sub)networks. Henceforth, we denote the 347-gene set as \underline{GTEx-DAG}. 
    
    \item GTEx \cite{jia2018analysis} contains 710 genes whose expressions increase with age (i.e., are \emph{up-regulated}), of which 161 are present in all eight considered (sub)networks. Henceforth, we denote the 161-gene set as \underline{GTEx-UAG}.
    
    \item Lu \emph{et al.} \cite{lu2004gene} identified 442 genes whose expressions vary with age, of which 265 are present in all eight considered (sub)networks. 
    
    \item Berchtold \emph{et al.} \cite{berchtold2008gene} identified 8,277 genes whose expressions vary with age, of which 2,312 are present in all eight considered (sub)networks. 
    
    \item Simpson \emph{et al.} \cite{simpson2011microarray} identified 2,911  genes whose expressions vary across different stages of Alzheimer's disease, of which 930 genes are present in all eight considered (sub)networks.  
    
\end{itemize}

\subsubsection{Two considered definitions of (non-)aging-related gene labels\label{sect:method-aging}}

For supervised classification (discussed in the following sections), we need a set of genes labeled as aging-related (positive class) and a set of genes labeled as non-aging-related (negative class). Because GenAge is considered to be one of the most confident aging-related ground truth data sources, our primary label definition is with respect to GenAge. Human genes in GenAge are sequence orthologs of aging-related genes in model organisms. However, human aging is a complex process that has unique biological aspects compared to the aging process in a model species \cite{danchin2018bacteria}, which might be missed by GenAge. So, to hopefully capture the human-unique aspects, we separately consider a secondary label definition with respect to GTEx \cite{jia2018analysis}, which was obtained by studying aging directly in human. Specifically, of the entire GTEx, we focus on GTEx-DAG, because we analyze PPI data, and because genes in GTEx-DAG were shown to be relevant for PPIs, unlike genes in GTEx-UAG \cite{jia2018analysis}. The two label definitions, i.e., based on either GenAge or GTEx-DAG, are as follows.

\noindent\textbf{\emph{Primary definition with respect to GenAge.}} We label  as \underline{aging-related} those 185 genes from GenAge that exist in all eight considered (sub)networks. Because we want our non-aging-related gene set to be as confident as possible, we label as \underline{non-aging-related} those genes that exist in all eight considered (sub)networks but not in any of the six considered aging-related ground truth data; there are 1,485 such genes. We intentionally use the same 185-gene positive class and the same 1,485-gene negative class for classification in all eight (sub)networks. This allows us to fairly compare classification accuracy (i.e., accuracy of aging-related gene prediction) across all (sub)networks.

\noindent\textbf{\emph{Secondary definition with respect to GTEx-DAG.}} We label as \underline{aging-related} those 347 genes from GTEx-DAG that exist in all eight (sub)networks. We label as \underline{non-aging-related} those 1,485 genes that exist in all eight (sub)networks but not in any of the six considered aging-related ground truth data. Note that we use the GenAge definition separately from the GTEx-DAG definition, i.e., we perform classification for each of the two individually. This is because the two are very different (sequence- vs. expression-based) data that possibly capture different aspects of the aging process. Indeed, of the 185 GenAge-based  and 347 GTEx-based aging-related genes, only 17 genes are common to the two gene sets.

\subsubsection{Cancer-related genes} \label{subsect:cancerdata}

Vogelstein \emph{et al.} \cite{vogelstein2013cancer} identified 138 mutation-related cancer driver genes, i.e., genes whose intragenic mutations contribute to cancer. Sondka \emph{et al.} \cite{sondka2018cosmic} identified 723 cancer driver genes by manually curating cancer-related genes from COSMIC to determine the presence of somatic mutation patterns in cancer. We combine these two sets of cancer-related genes, and obtain 728 unique cancer-related genes. Of these 728 genes, 366 are present in each of the eight (sub)networks analyzed in these study, which we denote as \underline{cancer-related genes}. Of these 366 genes, 49 are aging-related and 87 are non-aging-related according to the primary GenAge-based definition. Also, among these 366 genes, 15 are aging-related and 87 are non-aging-related according to the secondary GTEx-DAG-based definition. 

\subsection{Predictive models: node features and classifiers} \label{sect:methods_predmodel}

\subsubsection{The considered existing unweighted dynamic features}

\begin{itemize}
    \item \underline{DGDV}, dynamic graphlet degree vector \cite{hulovatyy2015exploring}, counts how many times a node participates in each considered dynamic graphlet. Dynamic graphlets are an extension of static graphlets (small subgraphs on up to $n$ nodes) to the dynamic context, where the temporal order in which events occur in a dynamic network is added onto edges of a static graphlet. As in the original DGDV paper \cite{hulovatyy2015exploring}, we use up to 4-node and 6-event graphlets, resulting in a 3,727-dimensional DGDV node feature.
    
    \item \underline{GoT}, graphlet orbit transitions \cite{aparicio2018graphlet}, of a node counts, for every pair of considered graphlets, how many times one graphlet changes into the other. As in the original GoT publication \cite{aparicio2018graphlet}, and for a fair comparison with DGDV, we use 4-node graphlets, resulting in a 121-dimensional GoT node feature. 
    
    \item \underline{GDC}, graphlet degree centrality  \cite{milenkovic2011dominating}, is defined for a  node in a static network (as discussed below). We use GDC to compute, for each node $v$, $v$'s centrality in each of the 37 snapshots of a dynamic subnetwork. Then, we combine $v$'s 37 GDC values into its 37-dimensional dynamic GDC feature. We do the same for all other considered centrality features (listed below), each of which results in a 37-dimensional node feature. Going back to the definition of GDC in a static network: it ranks a node $v$ as central if $v$ participates in many graphlets or in complex (large or dense) graphlets (or both). 
   
    \hspace{0.5cm}For GDC, we use the code by Faisal and Milenkovi\'{c} \cite{faisal2014dynamic}, which considers up to 5-node graphlets without an option to customize the graphlet size. 
    
    \item \underline{ECC}, eccentricity centrality  \cite{faisal2014dynamic}, of a node $v$ calculates distance (i.e., the shortest path length) between  $v$  and each other node in the network, finds a node $u$ that is the most distant to $v$, and computes the reciprocal of the distance between $u$ and $v$. 

    \item \underline{KC}, $k$-coreness \cite{faisal2014dynamic}, of a node $v$ is the size of the largest network core to which $v$ belongs. A network core is a subgraph in which each node is connected to at least $k$ other nodes. When $v$ belongs to a $k$-core, it also belongs to 1-core, 2-core, 3-core, ...,  ($k$-1)-core. Then, KC of $v$ is the size of its largest $k$-core.  
    \item \underline{DegC}, degree centrality \cite{faisal2014dynamic}, of a node $v$ measures how many other nodes in the network $v$ is connected to. 
    
    \item \underline{CentraMV}, centrality mean and variation \cite{li2019supervisedBIBM,li2020supervisedTCBB}, of a node $v$ measures, for each of the four considered centrality-based features (GDC, ECC, KC, and DegC), two quantities: the mean and variation over $v$'s 37 centrality values. These two quantities for each of the four centrality-based features combined form an 8-dimensional CentraMV node feature.
\end{itemize}

\subsubsection{The considered existing unweighted static features\label{method:stat-feature}}

\begin{itemize}
    \item \underline{SGDV}, static graphlet degree vector \cite{milenkovic2008uncovering}, a static counterpart of DGDV, counts how many times a node participates in each considered static graphlet. Just as for DGDV and GoT, we consider up to 4-node static graphlets. This results in a 15-dimensional SGDV node feature. We have tested up to 5-node SGDV in this study, but this has not yielded an improvement (results not shown). So, for simplicity, we just report results for up to 4-node SGDV.
    
    \item \underline{UniNet} \cite{kerepesi2018prediction}  aggregates a number of node centralities (DegC, ECC, KC, average shortest path, betweenness, closeness, clustering coefficient, neighborhood connectivity, radiality, stress, and topological coefficient) into an 11-dimensional node feature. 
    
    \item \underline{$m$BPIs} \cite{freitas2011data} works as follows. First, we rank the genes (i.e., nodes) in the network based on their degrees from high to low, where nodes with the same degree are ranked the same. Then, we take the $m$ highest-degree nodes from the ranked list of nodes. Then, for a node $v$, its feature has $m$ dimensions corresponding to the $m$ nodes, where each dimension $d$ of the node $v$'s feature has a value of 1 if $v$ interacts with the top $m$ highest-degree node $d$, or it has a value of 0 if $v$ does not interact with $d$. Just as \cite{li2019supervisedBIBM,li2020supervisedTCBB}, and as suggested in the original $m$BPIs publication, we use $m = 30$ in this paper. This results in a 30-dimensional $m$BPIs node feature. Two exceptions are for NetWalk-Static and NetWalk-Static*, where we use $m$BPIs node features with a dimension of 31 instead of 30. This is because in each of the subnetworks, two genes are tied at rank 30, and so we include both genes in the set of top 30 highest-degree nodes. 
    
\end{itemize}

\subsubsection{The considered existing weighted dynamic features\label{method:wDynafeatures}}

\begin{itemize}
    \item \underline{DegC-wt} \cite{opsahl2010}, weighted degree centrality (or strength), of a node is defined as the sum of the weights of edges connecting the node to its direct neighbors.
  
    \item \underline{ClusC-wt} \cite{Barrat2004}, weighted clustering coefficient captures how well connected are the direct neighbors of a node, while also taking into account weights of the edges connecting the node to its direct neighbors and the edges among direct neighbors. Several definitions of weighted clustering coefficient of a node has been defined. Because it has been shown that there is no best weighted clustering coefficient definition \cite{jari2007}, given a node $i$, we use the following definition by Barrat \emph{et al.} \cite{Barrat2004}: $C(i) = \frac{1}{s_i\cdot(d_i-1)}\sum_{j,k}\frac{W_{ij}+W_{ik}}{2}$. Here, $W_{ij}$ and $W_{ik}$ are the weights of edges connecting the node $i$ to nodes $j$ and $k$, respectively, where $j$ and $k$ should be connected to each other, $s_i$ is the strength of the node $i$, and $d_i$ is the degree of the node $i$. 
    
    \item \underline{CloseC-wt} \cite{opsahl2010}, weighted closeness centrality measures how  close  a  node  is  on  average  to  all  other  network nodes, by computing the inverse of the sum of weighted shortest paths from the given node to all of the other network nodes. Here, a weighted shortest path between any two nodes is the path whose sum of the inverse of the corresponding edge weights is minimum.
    
    \item \underline{BetwC-wt} \cite{opsahl2010}, weighted betweenness centrality of a node measures the fraction of all weighted shortest paths in a network that pass through the given node. Here, a weighted shortest path is defined in the same way as above.
    
    \item \underline{EigenC-wt} \cite{newman2004}, weighted eigen vector centrality  measures the importance of a node based on how often the node is connected to other nodes of higher weighted degrees. More formally, weighted eigen vector centralities of nodes in a network correspond to the values in the first eigen vector of the weighted adjacency matrix of the corresponding network.
\end{itemize}

\subsubsection{The proposed weighted features\label{method:wdyna-feature}}

Different types of local neighborhoods of a node typically capture different characteristics of the node's network position. Hence, given a weighted network, we propose to extract four different features of a node that are based on four different network neighborhood types. That is, for a given node $v$, we extract its features based on \emph{(i)} edges connecting $v$ to its direct (one-hop) neighbors, \emph{(ii)} edges among one-hop neighbors of $v$, \emph{(iii)} edges between one-hop neighbors of $v$ and two-hop neighbors of $v$, and \emph{(iv)} edges among two-hop neighbors of $v$. We characterize a neighborhood type of a node by measuring the distribution of edge weights in the neighborhood type, i.e., for each edge weight, we count how many times the given edge weight occurs in the given neighborhood type. Consequently, given a dynamic weighted subnetwork (i.e., $w$NetWalk-Dynamic) with multiple snapshots, we create a feature vector of a node using three different approaches, as follows.

\begin{itemize}
  \item Approach 1: using weight distributions. For a given node, for a given neighborhood type, we compute weight counts (i.e., distribution of weights) for each snapshot separately. Then, we concatenate the weight counts from each of the snapshots into a single feature vector. This results in four feature vectors, one corresponding to each of the four neighborhood types. Additionally, we define a feature vector that first concatenates feature vectors from each of the four neighborhood types for each of the snapshots separately, and then combines these concatenated feature vectors from each of the snapshots into a single feature vector. So, in total, we create five feature vectors for each node.
  
  \item Approach 2: using correlation of weight distributions. For a given node, for a given neighborhood type, we measure the Pearson correlation of weight counts between each pair of snapshots, which results in a correlation matrix of size $N \times N$, where $N$ is the number of snapshots. We take the upper triangular matrix of this correlation matrix as a feature vector. We do this for each of the four neighborhood types separately, which results in four features. Additionally, we concatenate the feature vectors corresponding to the four neighborhood types to obtain another feature vector. So, in total, we create five features for each node.
  
  \item Approach 3: using comparison of weight distributions to a background weight distribution. For a given node, for a given neighborhood type, and for a given snapshot, we measure the deviation of the distribution of the neighborhood weights from the distribution of weights in the whole weighted dynamic subnetwork. We use the Cram\'{e}r-von Mises statistic \cite{anderson1962} to measure the deviation between two weight distributions. This gives a single number that captures the distance between the two distributions. Then, given $N$ snapshots, we combine the $N$ values to obtain a single feature vector. We do this for each of the four neighborhood types separately, which results in four feature vectors. Additionally, we concatenate the feature vectors corresponding to the four neighborhood types of a node to obtain another feature vector. So, in total, we create five feature vectors for each node.
\end{itemize}

The above procedure results in $15$ weighted dynamic feature vectors for a node. In addition to the above way of characterising the distribution of weights, where we count the occurrences of ``raw'' weights, we characterize the distribution of ``binned'' weights, as follows. We first divide the interval between the smallest and the largest weights in a weighted dynamic subnetwork into equal-sized bins of size of 1, and then we assign all of the weights in the weighted dynamic subnetwork to their corresponding bins. Finally, we count the number of assigned weights in different bins, i.e., count how many weights belong to a given bin, to obtain the corresponding weight distributions. Finally, we use the same three approaches as above, i.e., approach 1, approach 2, and approach 3, to create $15$ additional weighted dynamic feature vectors. Hence, in total, we create $30$ weighted dynamic feature vectors, $15$ without using raw (i.e., ``nobin'') weights and $15$ using binned (i.e., ``bin'') weights.

Given an aging-related ground truth data, we compare the $30$ weighted dynamic features with respect to their ability to predict aging-related genes, in order to select the best weighted dynamic feature. Given $w$NetWalk-Dynamic, for the primary definition of aging- and non-aging-related genes using GenAge data, the best weighted dynamic feature vector corresponds to the case when we use distributions of raw weights (i.e., approach 1 above) of the second neighborhood type of a node. We call this feature vector as ``Diff-nobin-2''. Given $w$NetWalk-Dynamic, for the secondary definition of aging- and non-aging-related genes using GTEx-DAG data, the best weighted dynamic feature vector corresponds to the case when we use distributions of binned weights (i.e., approach 1 above) of the first neighborhood type of a node. We call this feature vector as ``Diff-bin-1''.

To fairly compare $w$NetWalk-Dynamic to its static counterpart (i.e., $w$NetWalk-Static*), we create the corresponding static version of the best weighted dynamic feature, as follows. Given  $w$NetWalk-Static*, we use the same approach to extract a weighted static feature that we use to create the best selected weighted dynamic feature. We identify the corresponding static features as ``Static-nobin-2'' and ``Static-bin-1'' that correspond to Diff-nobin-2 and Diff-bin-1, respectively.

\subsubsection{Two considered feature dimensionality reduction choices} \label{sect:method-dimension}

A common problem in supervised classification is overfitting, which often happens due to high dimensions of features used in the classification task \cite{subramanian2013overfitting}. Some of our considered features, including DGDV, GoT, and our proposed weighted dynamic features, have high dimensions. So, we apply linear feature dimensionality reduction (i.e., PCA) to all of our features. We are aware of more recent proposed nonlinear dimensionality reduction techniques, such as $t$-distributed stochastic neighbor embedding ($t$SNE)\cite{maaten2008visualizing}. We tested $t$SNE under six perplexity parameters (5, 13, 21, 30, 40, 50) in our previous work \cite{li2020supervisedTCBB}. However, we did not find any such case in which $t$SNE performs the best in terms of prediction accuracy, which is why we no longer test $t$SNE in this study. Hence, for each feature,  we consider: \emph{(i)} the full feature (i.e., no dimensionality reduction) and \emph{(ii)} linear PCA that considers as few principal components as needed to account for at least 90\% of variation in the data corresponding to the given feature. We perform feature dimensionality reduction during cross-validation, same as our previous study \cite{li2020supervisedTCBB}.

\subsubsection{Three considered classifiers\label{sect:methods-classifiers}}

Among the nine classifiers that we considered in our previous studies \cite{li2019supervisedBIBM,li2020supervisedTCBB}, we select three classifiers that perform better than the other classifiers in supervised prediction of aging-related genes, i.e., LR \cite{yu2011dual}, SVM-rbf  \cite{fan2008liblinear}, and NB \cite{chan1982updating}. Briefly, LR is a classification algorithm that uses a logistic function to model a binary dependent variable. It usually performs well when the features are approximately linear and targets are linearly separable. NB is a family of simple ``probabilistic classifiers'' that applies the Bayes’s theorem with an assumption that the features are independent of each other. We use Gaussian NB in our study. SVM outputs an optimal hyperplane by using a hinge loss function to categorize objects. It can be used with multiple kernel functions depending on whether the objects are linearly separable or not. We use a non-linear kernel (radial basis function (\underline{SVM-rbf})). All classifiers are implemented using a Python library scikit-learn (version 0.23.1) \cite{scikit-learn}.

\subsection{Evaluation \label{sect:evaluation}}

All methodology in this section is described when using the primary GenAge-based definition of aging- and non-aging-related gene labels. Everything is analogous when using the secondary GTEx-DAG definition.

\subsubsection{Predictive models}

For each definition of aging- and non-aging-related gene labels, for each of the eight (sub)networks, we develop a set of \emph{predictive models}, each derived from unique combinations of a node feature, a dimensionality reduction choice, and a classifier. In particular, we consider 46 node features, where seven are unweighted dynamic, three are unweighted static, 35 are weighted dynamic, and one is weighted static. For each feature, we consider two dimensionality reduction choices (i.e., PCA vs no dimensionality reduction). Then for each feature version (i.e., PCA reduced version and full version), we consider three classifiers. Note that for weighted dynamic features, we only perform PCA on the best existing and the best proposed weighted dynamic feature because performing PCA on every weighted dynamic feature is computational expensive due to their high dimensionalities. If some of the weighted features perform even better under PCA compared to no dimensionality reduction, it would further strength our hypotheses. We summarize the number of predictive models (i.e., 234) in Table \ref{table:pred-models}. Note that these 234 models are just for the GenAge-based definition of aging- and non-aging-related gene labels. The number is the same for the GTEx-based definition. So, over the entire study, we develop and evaluate a total of 468 predictive models. We present this number to give an idea to the reader of how comprehensive our study is. 

\begin{table}[!htp]
    \centering
    \caption{The number of predictive models that we test for each (sub)network. In total, we consider $18+42+18+42+18+18+70+4 = 230$ models with respect to GenAge-based definition of aging- and non-aging-related gene labels. Furthermore, because we apply PCA only on the best existing and the best proposed weighted dynamic features, we further consider $2 \times 1 \times 2 = 4$ models, which totals to 234 predictive models. The number of considered predictive models for GTEx-DAG-based definition of aging- and non-aging-related gene labels is also 234. Thus, in total, we consider 468 predictive models in this study.}
    \label{table:pred-models}
    \begin{tabular}{|l|l|c|c|c|c|}
    \hline
    Network type & Network name  & \begin{tabular}[c]{@{}c@{}}Feature \\ count\end{tabular} & \begin{tabular}[c]{@{}c@{}}Dimensionality\\ choices\end{tabular} & \begin{tabular}[c]{@{}c@{}}Classifier\\ count\end{tabular} & \begin{tabular}[c]{@{}c@{}}Predictive \\model\\ count\end{tabular}   \\ \hline
    Unweighted static & Entire                    & 3   & 2  & 3   & 18  \\ \hline
    Unweighted dynamic & Induced-Dynamic          & 7   & 2  & 3   & 42  \\ \hline
    Unweighted static & Induced-Static            & 3   & 2  & 3   & 18  \\ \hline
    Unweighted dynamic & NetWalk-Dynamic          & 7   & 2  & 3   & 42  \\ \hline
    Unweighted static & NetWalk-Static            & 3   & 2  & 3   & 18  \\ \hline
    Unweighted static & NetWalk-Static*           & 3   & 2  & 3   & 18  \\ \hline
    Weighted dynamic & $w$NetWalk-Dynamic         & 35  & 1  & 2   & 70  \\ \hline
    Weighted static & $w$NetWalk-Static*          & 1   & 2  & 2   & 4   \\ \hline
    \end{tabular}
\end{table}

\subsubsection{Hyperparameter training, cross-validation and prediction of aging-related genes\label{sect:5-fold-cv}}

We run each predictive model under a systematic 5-fold cross-validation framework \cite{li2020supervisedTCBB} via two steps, \emph{(i)} hyperparameter tuning and \emph{(ii)} model training and testing.

Specifically, we first randomly split the 185 aging- and 1485 non-aging-related genes into five equal-size subsets and perform 5-fold cross-validation. For each fold, one subset of the aging- and non-aging-related genes is treated as the testing data, and the remaining four subsets are treated as the training data. We denote them as model testing and model training data. Unlike our previous study \cite{li2019supervisedBIBM,li2020supervisedTCBB}, where we used the default hyperparameter values for the considered classifiers, in our present study, we use each model training data to perform hyperparameter tuning, in order to give each model the best-case advantage. To do this, we further perform 5-fold cross-validation. That is, given a model training data, we randomly split the aging- and non-aging-related genes into five equal-size subsets. A subset of the aging- and non-aging-related genes is treated as the testing data, and the remaining four subsets are treated as the training data. We denote them as tuning testing and tuning training data because they are generated for hyperparameter tuning. We train a given model via different hyperparameters on tuning training data and test it on tuning testing data. We quantify the performance using the average AUPR and select the ``best'' set of hyperparameter(s) that yields the highest average AUPR over the 5 folds. Note that this is only for one fold of the model training and model testing data. Because we have five folds, we have five sets of hyperparameter(s) (each set corresponding to a fold) for a given predictive model. We perform hyperparameter tuning for both LR and SVM-rbf. There is no hyperparameter for NB, so we use the default setting. The hyperparameter for LR is the regularization strength for which we select 10 values between $2^{-8}$ to $2^8$. There are two hyperparameters for SVM-rbf, i.e., gamma and regularization strength. We select 10 values between $2^{-8}$ to $2^8$ for each of the two hyperparameters, which results in $10 \times 10 = 100$ sets of hyperparameters via grid search.

Given the best predictive model, we test the trained model on the testing data. Given a testing set, the output is a ranked list of the genes based on their predicted likelihood of being aging-related. If multiple genes have the same probability of being aging-related genes, their ranks are the same. For example, probabilities of $1.0, 0.95, 0.95, 0.9$ are ranked as $1, 2, 2, 4$. Then, to make aging-related gene predictions using a model, a threshold $k$ needs to be selected so that those genes whose likelihoods of being aging-related are above the threshold are predicted as aging-related. We vary the number of predictions $k$ from 1 to $\lceil (185+1485)/5 \rceil = 334$, in the increments of 1.

\subsubsection{Evaluation measures}

For each fold, for predicted genes at each prediction threshold $k$, we use their actual aging- and non-aging-related labels to compute the numbers of true positives (TPs), false positives (FPs), and false negatives (FNs). A gene is considered to be TP if it is predicted to be aging-related and is also labeled as aging-related. A gene is considered to be FP if it is predicted to be aging-related but is labeled as non-aging-related. A gene is considered to be FN if it is \emph{not} predicted to be aging-related but is labeled as aging-related. 

Given TPs, FPs, and FNs for each prediction threshold for each of the five folds, we evaluate accuracy of a predictive model using average precision, average recall, average F-score, and average AUPR over five folds, where, \underline{precision} is \# of TPs$/($\# of TPs $ + $ \# of FPs$)$, \underline{recall} is \# of TPs$/($\# of TPs $ + $ \# of FNs$)$, \underline{F-score} is the harmonic mean of precision and recall, and AUPR is the area under the precision recall curve. Specifically, we compute these measures as follows.

Given a predictive model, we compute average precision, average recall, and average F-score using the following procedure. First, we select a single prediction threshold $k1$, which is the same for all of the five folds. Second, for each fold, we compute fold-specific precision, recall, and F-score at $k1$. Third, we take the average of fold-specific precision, recall, and F-score values at $k1$ over all of the five folds to obtain the average precision, average recall, and average F-score, respectively. We select the prediction threshold $k1$ as follows. For each prediction threshold $k$, we combine the predictions from each fold into a single list, and calculate the precision, recall, and F-score.  We believe that for potential wet lab validation of predictions, precision should be favored over recall, as long as recall is not too low \cite{li2019supervisedBIBM,li2020supervisedTCBB}. Because maximizing F-score is exactly what ensures that recall is not too low, i.e., that both precision and recall are reasonably high, we select the prediction threshold $k1$ where F-score is maximized, while at the same time precision is at least as large as recall.

Given a predictive model, we compute average AUPR as follows. First, given precision and recall values for each prediction threshold $k$ in a given fold, we compute the fold-specific AUPR. Then, we take average of the fold-specific AUPRs over the five folds to obtain a single average AUPR value.

Note that after we compute the above average precision, average recall, average F-score, and average AUPR values, we round each of them to the second decimal place to avoid marginal superiority of one predictive model to another. Specifically, considering accuracy values rounded to second decimal place ensures that we only consider a  predictive model to have higher accuracy than another if the former is at least $0.5\%$ more accurate than the latter.

\subsubsection{Choice of the best predictive model for each (sub)network}

For each of the eight (sub)networks, we rank the predictive models based on their average AUPR, average F-score, average precision, and average recall. Then, we select the ``best'' predictive model with the highest rank, i.e., the one that maximizes the average AUPR, to give it the best-case advantage. Then, we report average AUPR, average precision, average recall, and average F-score of the selected predictive model.

\subsubsection{Statistical significance of a predictive model's  accuracy\label{sect:statistical}}

First, we compare accuracy of each selected predictive model against accuracy of a random approach. The latter works as follows. For a given testing data (i.e., cross-validation fold) and each prediction threshold $k$, we randomly select $k$ testing genes and predict them as aging-related. Then, we repeat this for each of the five folds. To account for randomness, we run these two steps 30 times, which means that we perform $5\times 30 = 150$ random aging-related gene predictions. We report the accuracy scores (AUPR, precision, recall, and F-score) of the random approach as the average over the 150 runs. A predictive model is good only if its accuracy is statistically significantly higher than that of a random approach.

Second, we compare pairs of the selected predictive models (plus the random approach), to determine if one model's predicted gene set is statistically significantly more accurate than another model's predicted gene set. We do this by comparing the two models' five accuracy scores corresponding to the five folds via the paired Wilcoxon signed-rank test \cite{wilcoxon1992individual}. Because we perform this test for multiple pairs of models, we apply the false discovery rate (FDR) correction \cite{benjamini1995controlling} to adjust the $p$-values. We choose $0.05$ as the significance level (i.e., one model is statistically significantly better than another if the adjusted $p$-value is below $0.05$).

\subsubsection{Measuring overlaps between prediction sets\label{sect:overlaps}}

To evaluate potential complementarity of two gene sets (e.g., $A$, $B$), we measure their overlap via the Jaccard index, $J(A, B) = \frac{A \cap B}{A \cup B}$. We compute the statistical significance of the overlap size using the hypergeometric test \cite{rivals2007enrichment}. Because we run this test for multiple pairs of predictive models, we correct the $p$-values using the FDR correction. We choose $0.05$ as the significance level (i.e., the overlap is statistically significantly large if the adjusted $p$-value is below $0.05$).


\section*{\small Competing interests}
\footnotesize The authors declare that they have no competing interests.
 
\section*{\small Funding}
\footnotesize This work including publication costs is funded by the U.S. National Science Foundation (NSF) CAREER award CCF-1452795. The funding body had no role in the design of the study and collection, analysis, and interpretation of data and in writing the manuscript.

\section*{\small Author's contributions}
\footnotesize Conceptualization: Q.L., K.N. and T.M. Data collection and process: Q.L. and K.N. Methodology and experiments: Q.L. and K.N. Results visualization and analysis: Q.L. and K.N. Writing: Q.L., K.N. and T.M. Supervision: T.M. Project administration: T.M. Funding acquisition: T.M. All authors have read and approved the manuscript.

\bibliographystyle{unsrt}  
\bibliography{references}


\end{document}